\documentclass[amssymb,amsmath,superscriptaddress,twocolumn,footinbib]{revtex4}
\usepackage{natbib,graphicx,subfigure,subeqnarray,fancyhdr,amsmath,epstopdf,multirow,color, mathrsfs}
\begin{document}

\newcommand{\change}[1]{{\color{black} #1}}
\renewcommand{\labelitemi}{$-$}
\newcommand{\Sc}{\mathcal{S}}
\newcommand{\Fc}{\mathcal{F}}\newcommand{\Rc}{\mathcal{R}}\newcommand{\dd}{\mathrm{d}}
\newcommand{\ee}{\mathrm{e}}\newcommand{\ci}{\mathrm{i}}\newcommand{\ib}{\mathbf{i}}
\newcommand{\jb}{\mathbf{j}}\newcommand{\kb}{\mathbf{k}}\newcommand{\ab}{\mathbf{a}}
\newcommand{\Fb}{\mathbf{F}}\newcommand{\fb}{\mathbf{f}}\newcommand{\Gb}{\mathbf{G}}
\newcommand{\Mb}{\mathbf{M}Ä}\newcommand{\nb}{\mathbf{n}}\newcommand{\Sb}{\mathbf{S}}
\newcommand{\Sbs}{\mathbf{S^*}}\newcommand{\Rb}{\mathbf{R}}\newcommand{\Sigb}{\boldsymbol{\Sigma}}
\newcommand{\Sigbs}{\boldsymbol{\Sigma^*}}
\newcommand{\omegab}{\boldsymbol{\omega}}\newcommand{\betab}{\boldsymbol{\beta}}
\newcommand{\alphab}{\boldsymbol{\alpha}}
\newcommand{\epsb}{\boldsymbol{\epsilon}}
\newcommand{\ub}{\mathbf{u}}
\newcommand{\eb}{\mathbf{e}}\newcommand{\vv}[1]{\underline{#1}}\newcommand{\ev}{\vv{e}}
\newcommand{\rv}{\vv{r}}\newcommand{\TT}[1]{\underline{\underline{#1}}}\newcommand{\omb}{\mathbf{\omega}}
\newcommand{\Ub}{\mathbf{U}}\newcommand{\xb}{\mathbf{x}}\newcommand{\rb}{\mathbf{r}}
\newcommand{\ssb}{\mathbf{s}}\newcommand{\Xb}{\mathbf{X}}
\newcommand{\Rey}{{\rm Re}}
\newcommand{\Pe}{\mbox{Pe}\,}
\newcommand{\Pee}{\mbox{Pe}}
\newcommand{\mean}[1]{\langle #1\rangle}
\newcommand{\ddp}{[p]^\pm}\newcommand{\taub}{\mbox{\boldmath$\tau$}}\newcommand{\Fr}{\mbox{\textit{Fr}}}
\let\grad\nabla\newcommand{\z}{\zeta}\newcommand{\kk}{\kappa}\newcommand{\tkk}{\tilde{\kappa}}
\newcommand{\e}{\varepsilon}\newcommand{\zb}{\bar{\zeta}}\let\grad\nabla\let\bcdot\cdot
\newcommand{\half}{{\textstyle\frac{1}{2}}}
\newcommand{\textfrac}[2]{{\textstyle\frac{#1}{#2}}}
\newcommand{\LF}[1]{{#1}^{\mathrm{LF}}}\newcommand{\Lap}[1]{{#1}^{\mathrm{L}}}
\newcommand{\ds}{*\!*}\newcommand{\cond}[2]{\frac{\mathrm{D} #1}{\mathrm{D} #2}}
\newcommand{\pard}[2]{\frac{\partial #1}{\partial #2}}\newcommand{\totd}[2]{\frac{\mathrm{d}#1}{\mathrm{d}#2}}
\newcommand{\pardd}[3]{\frac{\partial^2 #1}{\partial #2 \partial #3}}
\newcommand{\Real}{\mbox{Re}}\newcommand{\Imag}{\mbox{Im}}
\newcommand{\Fpint}{=\!\!\!\!\!\!\!\int}
\newcommand{\txi}{\tilde\xi}\newcommand{\dxi}{\delta\xi}
\newcommand{\tpsi}{\tilde\psi}\newcommand{\dpsi}{\delta\psi}
\newcommand{\cin}{\mathscr{C}}
\newcommand{\gin}{\mathscr{G}}
\makeatletter
\def\sgn{\mathop{\operator@font sgn}}
\makeatother

\title{Locomotion as an optimal steady feeding mechanism for all P\'eclet numbers}
\title{Optimal feeding by steady swimming for all P\'eclet numbers}
\title{Optimal feeding is optimal swimming}
\title{Optimal feeding is optimal swimming for all P\'eclet numbers}
\author{S\'ebastien Michelin}
\email{sebastien.michelin@ladhyx.polytechnique.fr}
\affiliation{LadHyX -- D\'epartement de M\'ecanique, Ecole
  polytechnique, 91128 Palaiseau Cedex, France.}
\author{Eric Lauga}
\email{elauga@ucsd.edu}
\affiliation{Department of Mechanical and Aerospace Engineering, University of California San Diego, 9500 Gilman Drive, La Jolla CA 92093-0411, USA.}

\date{\today}
\begin{abstract}

Cells swimming in viscous fluids create flow fields which influence the transport of relevant nutrients, and therefore their feeding rate. We propose a modeling approach to the problem of optimal feeding at zero Reynolds number. We consider a simplified spherical swimmer deforming its shape tangentially  in a steady fashion (so-called  squirmer). Assuming that the nutrient is a passive scalar obeying an advection-diffusion equation, the optimal use of flow fields by the swimmer for feeding is determined by maximizing the diffusive flux at the organism surface   for a fixed rate of energy dissipation in the fluid. The results are obtained through the use of an adjoint-based numerical optimization implemented by a Legendre polynomial spectral method. We show that, to within a negligible amount,  the optimal feeding mechanism consists in putting all the energy expended by surface distortion  into  swimming -- so-called treadmill motion -- which is also the solution maximizing the swimming efficiency. Surprisingly, although the rate of feeding depends strongly on the value of the P\'eclet number, the optimal feeding stroke is shown to be essentially  independent of it, which is confirmed  by asymptotic analysis. Within the context of steady actuation, optimal feeding is therefore found to be equivalent to optimal swimming for all  P\'eclet numbers.

\end{abstract}
\maketitle

\section{Introduction}
\label{sec:intro}

Swimming microorganisms can be found in a variety of environments, and encompass a wide range of size and locomotion mechanisms \cite{lighthill1975,lauga2009}. For bacteria, motility is important to achieve many biological functions, including location and migration toward regions rich in nutrients, oxygen or light  \cite{fenchel2002,stocker2008},  swimming against gravity, or    escaping aggressions \cite{crawford1992,hamel2011}. Motility is also essential to  reproductive success, in particular for mammals \cite{suarez06}. Recently, the collective motion of dense swimmer suspensions was the focus of a number of studies emphasizing  instabilities and  increased  mixing  \cite{batchelor1970,pedley1992,hernandez2005,saintillan2007,saintillan2008b}. In order to swim in a viscous fluid, a microorganism must undergo sequences of active and non-time-reversible deformations of its body surface \cite{purcell1977,lauga2011}. This surface deformation sequence will be referred to in the following as the stroke, which could be either a swimming stroke (leading to a net displacement of the swimmer center of mass), or non-swimming. 

As such a swimmer performs work against the surrounding fluid, it creates a flow field and can thus modify its immediate  environment in an important fashion, affecting in particular the transport of nutrients.   The metabolism of many microorganisms relies on the absorption at their surface of various  particles or molecules  which are both diffusing  and  being advected by the swimmer-induced flow. Depending on the organism considered, these can range from dissolved gases or low-weight molecules, to complex proteins, organic compounds, small particles, or even sometimes heat. This is true from the behavior of small bacteria all the way  to large organisms such as the protozoon \emph{Paramecium} which feeds on smaller bacteria, whose typical random walk motion is equivalent to a diffusive process at the scale of the larger organism \cite{lovely1975,berg1993,garcia2011}. For simplicity, all these cases \change{will be referred to} as ``nutrients''.

An interesting transport problem in the dynamics of swimming cells concerns the coupling between the flow created by the swimmer and the transport of nutrients. \change{This coupling can be essential for larger cells or cell colonies to achieve feeding rates matching their metabolic needs \cite{short2006}.}
If $\kappa$ \change{is} the \change{diffusivity} of the nutrient of interest, and $a$ the typical scale of the organism, 
the impact of the stroke on  feeding is  characterized by the 
value of the  P\'eclet number,  $\Pe=\tau_\textrm{def}/\tau_\textrm{diff}$, where $\tau_\textrm{def}$ is the characteristic time scale for the shape deformation (stroke) and $\tau_\textrm{diff}=a^2/\kappa$ is the diffusive time scale around the organism. 

At small P\'eclet number, the concentration gradients created by the stroke-induced flow are immediately homogenized by diffusion, and therefore shape changes affect only marginally the instantaneous feeding rate. In that case, swimming can still affect  feeding indirectly by allowing to  access regions of higher nutrient concentration \cite{purcell1977}.  At large P\'eclet number, however,  the advective transport by  the flow created by the swimming stroke can significantly modify the nutrient concentration field. In that case, swimming directly impacts  feeding both by creating large concentration gradients near the body surface and by increasing the swimmer ability to scan a large volume of fluid \cite{childress1987}. 

The purpose of the present paper is to quantify the impact of the swimming stroke on the feeding ability of an organism and to determine the optimal stroke maximizing the nutrient uptake. A priori, the optimal stroke should   depend on the problem of interest   through the value  of the P\'eclet number. In Nature, the relevant value of $\Pee$ varies by several orders of magnitude, due to the large variety of sizes and time scales observed for different microorganisms (from less than $1\,\mu$m for the smallest bacteria to several hundred $\mu$m for larger eukaryotes) and  the range of diffusivity coefficients for the nutrients of interest (in aqueous solutions, $\kappa_T\sim 10^{-7}\,$m$^2\,$s$^{-1}$ for heat, $\kappa_{O_2}\sim 3\,10^{-9}\,$m$^2\,$s$^{-1}$ for oxygen and small molecules, and $\kappa\sim 10^{-11}$--$10^{-10}\,$m$^2\,$s$^{-1}$ for larger proteins). For a given organism, the optimal stroke  to maximize, for example,  heat fluxes  might therefore not be the same as the one maximizing the absorption of a large protein.

Performing the swimming stroke has an energetic cost for the organism. In this paper we will consider the portion of the energy budget which includes the rate of working against the fluid, which is instantaneously dissipated in the form of heat in the fluid.  The organism's metabolism imposes a restriction on the maximum energy available for motility, and assuming that energy losses other than hydrodynamic  can be embedded in a fixed metabolic efficiency, optimizing the swimming stroke for feeding is a mathematical problem which can be formulated as follows:  For a given amount of energy available to a particular microorganism to create a flow, what is the optimal  stroke (possibly a non-swimming one) that maximizes the nutrient uptake? 

In Stokes flow, both body and fluid inertia are negligible compared to viscous forces \cite{lauga2009}. The displacement of the microorganism and the hydrodynamic efficiency are then entirely determined by the shape change sequence and not by the rate at which this sequence is performed. Solving for the stroke-induced swimming motion and the corresponding velocity field around the organism can be tedious for complex geometries, as it generally involves the flapping motion of a few or many flexible flagella or cilia \cite{lighthill1975}.  Most of the available literature focuses on two possible modeling approaches. In the first one, each flexible appendage or body element is modeled individually using slender body theory \cite{lighthill1975,keller1976b} or singularity methods \cite{chwang1975}. The second approach, to which this work belongs, considers a simplified geometry for which the Stokes equations can be solved exactly. This is the case for the classical spherical squirmer model considered here \cite{lighthill1952,blake1971}, an envelope model for the dynamics of ciliated microorganisms that has been used previously to study hydrodynamics interactions \cite{ishikawa2006}, suspension dynamics \cite{ishikawa2007a,ishikawa2007b} and optimal locomotion \cite{michelin2010c}. For this model, the linearity of Stokes equations can be exploited to linearly decompose the stroke  in  a superposition of swimming and non-swimming modes, which can then be optimized to maximize the organism displacement for a given energetic cost \cite{leshansky2007,tam2007,michelin2010c,tam2011}. In recent work \cite{michelin2010c}, we showed that the optimal time-periodic swimming strokes, {\it i.e.}~the one leading to the largest swimming speed for a given amount of available viscous dissipation, exhibit wave patterns reminiscent of the metachronal waves observed on the surface of ciliated microorganisms \cite{brennen1977}. 

The effect of swimming on the transport of passive scalars has been studied in the past both from  Lagrangian and Eulerian points of view. In the Lagrangian approach, the capture or drift induced on a given particle by the  swimming motion of the organism is explicitly solved for \cite{childress1987,thiffeault2010,lin2011}. In the Eulerian approach, the organism is modeled as being suspended in a continuous concentration field of nutrients, and the focus is on the absorption flux on the swimmer body  \cite{magar2003,magar2005,tam2011}. The feeding of a model squirmer was recently addressed for steady and unsteady tangential surface motions described by the superposition of one swimming and one non-swimming mode \cite{magar2003,magar2005}. The nutrient uptake was observed to be strongly dependent on the  value of the P\'eclet number  as well as the relative intensity of the non-swimming and  swimming mode.

In the current paper we propose to determine the  optimal feeding  stroke for a squirmer, namely the one maximizing the uptake of a nutrient by the organism for a given hydrodynamic energetic cost. We consider the simplest swimmer geometry (a sphere) and focus, as our first attempt to solve the problem, on the case of a steady stroke where the imposed surface velocity is time-independent. Such an assumption is obviously  a simplification as cilia tips display periodic and  unsteady displacements.  As recently observed \cite{michelin2010c}, the optimal unsteady  stroke for locomotion can in fact be interpreted as the periodic regularization of the solution to the  steady optimal problem. It was also shown \cite{magar2005} that for some particular limit of large $\Pe$ and infinitesimal deformation, the average feeding by the unsteady stroke is defined at leading order by the result of a modified steady problem. 
Our determination of the optimal steady feeding stroke  is thus expected to provide   important physical insights on the relation between swimming and feeding for microorganisms. In addition, although results are presented here for an idealized organism shape, the optimization framework detailed in this paper is applicable to more complex geometries and is therefore relevant to a wide class of advection-diffusion problems near self-propelled organisms.

In this steady framework, the problem at stake is the optimal distribution of the available hydrodynamic energy between the different actuation modes of the swimmer,  either swimming modes that produce locomotion or non-swimming modes that only produce stirring of the surrounding fluid. To answer this question, the general framework of the steady feeding problem \change{is presented} in Sec.~\ref{sec:feeding_problem}. After considering an organism of arbitrary shape, the equations are \change{introduced} for the particular case of the squirmer and solved numerically for some specific strokes using a spectral method, allowing us to gain qualitative understanding of the effect of the swimming stroke on the concentration field and nutrient uptake. 
In Sec.~\ref{sec:optimal_feeding}, \change{we derive} an adjoint-based optimization procedure  to determine the optimal stroke for a general swimmer, and we apply it  to  \change{characterize} computationally the optimal stroke for the squirmer as a function of the  P\'eclet number. We show that, to within a negligible quantitative difference, optimal feeding is equivalent to optimal swimming for all P\'eclet numbers.  \change{Our numerical results are compared successfully to predictions of asymptotic analysis, at both large and small P\'eclet numbers. Finally, we close by a discussion in  Sec.~\ref{sec:discussion}.}

\section{Nutrient transport around a swimming microorganism}\label{sec:feeding_problem}
\subsection{Advection-diffusion of a passive scalar near a general swimming microorganism}
\label{sec:general_feeding}

We consider the transport of a passive scalar field  around a  microorganism which stirs the surrounding fluid -- and possibly swims as well --   by imposing a steady  tangential velocity along its surface, described by $\ub^{S}$. The surface $\Sc$ and the shape of the organism is  therefore assumed to remain  independent of time.  Throughout this paper, a body-fixed reference frame \change{is considered}. The Reynolds number, $\Rey=\rho Ua/\mu$ is assumed to be small, where $U$ and $a$ are the typical swimming velocity and length scale of the swimmer, and $\rho$ and $\mu$ are the density and dynamic viscosity of the fluid medium. For $\Rey\ll 1$, fluid and solid inertia can be neglected and the velocity field $\ub$ around the swimmer is solution of the incompressible Stokes problem
\begin{align}
 -\grad p+\mu\grad^2&\ub ={\bf 0},\label{eq:stokes}\\
\nabla\cdot\ub&=0,\\
\ub=\ub^{S}&\textrm{   \,for\,   }\xb\in\Sc,\\
\ub\rightarrow -(\Ub+\boldsymbol\Omega\times\xb) &\textrm{\,  for\,  }\xb\rightarrow\infty.\label{kine}
\end{align}
In Eq.~\eqref{kine}, the translation and rotation velocities, $\Ub$ and $\boldsymbol\Omega$, define the organism swimming motion and are determined by imposing the free-swimming conditions of zero net hydrodynamic force and torque \cite{childress1981}
\begin{subeqnarray}\label{eq:swimming}
\int_\Sc\boldsymbol\sigma\cdot\nb\,\dd S&=&0,\\
 \int_\Sc\xb\times(\boldsymbol\sigma\cdot\nb)\dd S&=&0,
\end{subeqnarray}
where $\boldsymbol\sigma=-p\mathbf{I}+\mu(\grad\ub+\grad^T\ub)$ is the stress tensor in the fluid, and $\nb$ the unit normal vector pointing into the fluid. The steady swimming problem in Eqs.~\eqref{eq:stokes}--\eqref{eq:swimming} is linear with respect to $\ub^S$ and its solution for the swimming velocities and fluid velocity fields can therefore be rewritten formally  as 
\begin{equation}
(\mathbf{U},\boldsymbol\Omega)=\mathscr{L}\cdot\ub^S,\qquad \ub=\mathcal{L}\cdot\ub^S\label{eq:lin_op},
\end{equation}
where $\mathscr{L}$  and $\mathcal{L}$ are linear operators depending solely on the swimmer geometry. 

The hydrodynamic  cost of the swimming motion, $\mathcal{P}$, is defined as the rate of work performed by the swimmer surface against the fluid, and equal to the energy dissipation rate by viscous stresses in the entire fluid domain $V_f$
\begin{equation}
\mathcal{P}=\int_{V_f}(\boldsymbol\sigma:\mathbf{d})\, \dd V=-\int_\Sc\ub^S\cdot(\boldsymbol\sigma\cdot\nb)\dd S,
\end{equation}
where $\mathbf{d}=(\grad\ub+\grad\ub^T)/2$ is the fluid strain rate tensor. In the following, the equations are non-dimensionalized using $a$ and $\sqrt{\mathcal{P}/\mu a}$ as reference length and velocity scales respectively.

The microorganism is assumed to be suspended in an unbounded nutrient solution with concentration $C=C_\infty$ in the far-field. The nutrient is assumed to be  totally absorbed by diffusion through the swimmer surface, and $C=0$ \change{is imposed} on $\Sc$. \change{Note that for a real swimmer, this assumption is only valid if the nutrient flux at the surface is smaller than the cell's metabolic processing rate (see Ref.~\cite{magar2003} for a discussion of a more realistic boundary condition).}

For convenience, the nutrient concentration is rescaled as $c=(C_\infty-C)/C_\infty$. With this rescaling, $c\ll 1$ corresponds to near-ambient nutrient concentrations in the far-field, while $c\sim 1$ corresponds to nutrient-depleted regions near the organism. The rescaled concentration field $c(\xb)$ is the solution of the steady advection-diffusion problem
\begin{align}
&\Pe\ub\cdot\grad c=\nabla^2 c\label{eq:nutrient},\\
&c=1 \textrm{ \,  for\,   } \xb\in\Sc,\label{eq:nutrient_bc1}\\
& c\rightarrow 0 \textrm{ \, for \, }\xb\rightarrow\infty,\label{eq:nutrient_bc2}
\end{align}
where $\ub$ is the velocity field solution of the swimming problem in Eqs.~\eqref{eq:stokes}--\eqref{eq:swimming}, and 
\begin{equation}\label{eq:peclet}
\Pe=\frac{1}{\kappa}\sqrt{\frac{\mathcal{P}a}{\mu}},
\end{equation}
is the P\'eclet number defined using the characteristic length and velocity scales and the  nutrient diffusion constant, $\kappa$. 

The flux of nutrient \change{on} the swimmer's surface is purely diffusive and thus defined in non-dimensional form as
\begin{equation}\label{eq:flux_gen}
\Phi=-\frac{1}{\Pe}\int_\Sc\pard{c}{n}\dd S,
\end{equation}
where $\partial c/\partial n=\nb\cdot\grad c$ and $\nb$ is the normal unit vector to the solid boundary pointing into the fluid domain. 
When $\ub^S=0$, the organism \change{(a rigid body) does not create any flow field and the energy consumption is $\mathcal{P}=0$. Then} $\Pe=0$, and the nutrient uptake is the solution to the purely diffusive problem in Eqs.~\eqref{eq:nutrient}--\eqref{eq:nutrient_bc2}, \change{ with a corresponding nutrient flux $\Phi_0$.}  Rather than the absolute nutrient uptake \change{$\Phi$ resulting from a given stroke}, we are interested here in its increase relative to the rigid body reference case, namely $J=\Phi/\Phi_0=\textrm{Sh}/2$, where $\textrm{Sh}$ is known as the Sherwood number \cite{magar2003}. 

The problem solved in this paper can be formulated as follows. For a given amount of energy available to the organism to stir the fluid (measured in a dimensionless fashion by $\Pee$), what is the stroke ({\it i.e.}~the surface velocity field $\ub^S$) that maximizes the relative nutrient uptake $J$? Note that non-dimensionalizing the problem using the energy used by the organism rather than its swimming velocity allows for both swimming ($\Ub\neq 0$) as well as non-swimming strokes ($\Ub=0$).

\subsection{The squirmer model}

\begin{figure}
\begin{center}
\includegraphics[width=8.5cm]{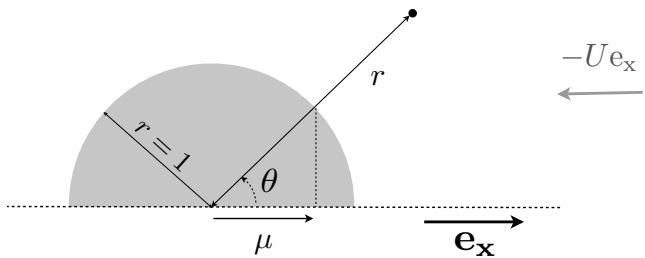}
\caption{Squirmer model and spherical polar coordinates used in the paper. On the surface of the swimmer ($r=1$), the fluid velocity is purely tangential $\ub=u_\theta^S(\mu)\eb_\theta$. In the far-field, $\ub\sim-U\eb_x$ with $U$ the swimming velocity of the organism.}\label{fig:squirmer}
\end{center}
\end{figure}

The general framework of the previous section is  now applied to the particular case of a spherical swimmer prescribing axisymmetric and steady surface velocities. By symmetry, the swimming motion of this so-called squirmer is at best a pure translation along a fixed direction $\eb_x$, and using spherical polar coordinates with respect to this axis  centered on the swimmer, all fields (velocities, pressure, nutrient concentration) only depend on $r$ and $\mu=\cos\theta$,  where $\theta$ is the polar angle with respect to $\eb_x$ (Figure \ref{fig:squirmer}). By taking $a$ to be the sphere radius, the swimmer  surface is the unit sphere $r=1$, and the surface velocity, $\ub^S=u^S_\theta\eb_\theta$, can be decomposed into  modes as \cite{blake1971}
\begin{equation}\label{eq:alphndef}
u^S_\theta(\mu)=\sum_{n=1}^\infty\alpha_nK_n(\mu),
\end{equation}
with
\begin{equation}\label{eq:Vndef}
K_n(\mu)=\frac{(2n+1)\sqrt{1-\mu^2}}{n(n+1)}\,L_n'(\mu),
\end{equation}
where $L_n(\mu)$ is the $n$-th Legendre polynomial. The swimming stroke is fully characterized by the values of the constant coefficients $\alpha_n$ ($n\geq 1$). The pressure $p$ and streamfunction $\psi$  can be computed at any point of the fluid domain as \cite{blake1971,michelin2010c}
\begin{eqnarray}
p(r,\mu)&=&p_\infty+\sum_{n=2}^\infty\alpha_nP_n(r,\mu),\\
P_n(r,\mu)&=&-\left(\frac{4n^2-1}{n+1}\right)\frac{L_n(\mu)}{r^{n+1}}\label{eq:p},
\end{eqnarray}
and
\begin{eqnarray}
\psi(r,\mu)&=&\sum_{n=1}^\infty\alpha_n\Psi_n(r,\mu),\label{eq:stream}\\
\Psi_n(r,\mu)&=&\frac{2n+1}{n(n+1)}(1-\mu^2)L_n'(\mu)\psi_n(r),\\
\psi_1(r)&=&\frac{1-r^3}{3r},\quad \psi_n(r)=\frac{1}{2}\left(\frac{1}{r^n}-\frac{1}{r^{n-2}}\right).\,\,\label{eq:stream_n}
\end{eqnarray}
The velocity field is easily recovered from $\psi$ as 
\begin{equation}
\ub=-\frac{1}{r^2}\pard{\psi}{\mu}\eb_r-\frac{1}{r\sqrt{1-\mu^2}}\pard{\psi}{r}\eb_\theta,\label{eq:velocity}
\end{equation}
and the swimming velocity is $U=\alpha_1$. Using this relation as well as Eqs.~\eqref{eq:stream}--\eqref{eq:velocity}, the linear operators $\mathcal{L}$ and $\mathscr{L}$ can be expressed in terms of $\mu$-projections on the Legendre polynomials.
 
In the squirmer model, the swimming stroke $\ub^S$ is thus entirely determined by the values of the different mode amplitudes $\alpha_n$, or equivalently the vector $\alphab$. \change{Note that the non-dimensonalization based on the rate of energy dissipation leads to the normalization} \cite{blake1971}
\begin{equation}\label{eq:energy_constraint}
\sum_{n=1}^\infty\beta_n^2=1,
\end{equation}
with $\beta_1=\alpha_1$, and
\begin{equation}\label{eq:beta}
\beta_n=\frac{(2n+1)\,\alpha_n}{\sqrt{3n(n+1)}}\textrm{ \,  for \,  }n\geq 2.
\end{equation}
With this rescaling, \change{all possible strokes correspond to a vector $\betab$ on the unit hypersphere} (in the remainder of the paper, \change{$\alphab$ and $\betab$ will be used equivalently }to characterize the swimming stroke). Note, that for mathematical convenience, the definition of the P\'eclet number in Eq.~\eqref{eq:peclet} was modified to 
\begin{equation}\label{eq:peclet2}
\Pe=\frac{1}{\kappa}\sqrt{\frac{\mathcal{P}a}{12\pi\mu}}\cdot
\end{equation}

The particular and so-called ``treadmill" squirmer must now be pointed out. That swimmer only includes one squirming mode ($\beta_n=\delta_{n1}$)  and maximizes the distance travelled by the swimmer for a given amount of energy \cite{leshansky2007,michelin2010c}.  For a general squirmer, the first mode ($n=1$) entirely defines the swimming velocity, and as such is referred to in the following as the swimming mode, as opposed to all the other modes ($n\neq 1$) which do not produce any swimming motion. The second mode ($n=2$) defines the  local stress  applied by the swimmer on the surrounding fluid \cite{batchelor1970, ishikawa2006}.  

For a given stroke $\alphab$, the rescaled nutrient concentration $c$ satisfies the advection-diffusion problem
\begin{subeqnarray}\label{eq:advdiff}
\Pe\sum_{n=1}^\infty\alpha_n\left[\pard{\Psi_n}{r}\pard{c}{\mu}-\pard{\Psi_n}{\mu}\pard{c}{r}\right]&=&\pard{}{r}\left(r^2\pard{c}{r}\right)\\
&&+\pard{}{\mu}\left((1-\mu^2)\pard{c}{\mu}\right),\nonumber\\
c(1,\mu)&=&1,\slabel{eq:advdiff_bc1}\\
 c(\infty,\mu)&=&0.\slabel{eq:advdiff_bc2}
\end{subeqnarray}
The reference nutrient flux $\Phi_0$ corresponds to the case of a non-stirring squirmer ({\it i.e.}~a rigid sphere \change{with} $\alpha_n=0$ for all $n$) for which the solution of Eqs.~\eqref{eq:advdiff} is simply $c_0=1/r$. From Eq.~\eqref{eq:flux_gen}, $\Phi_0=4\pi/\Pe$, and the relative nutrient uptake $J$ takes therefore the simple form
\begin{equation}
J=-\frac{1}{2}\int_{-1}^1\pard{c}{r}(1,\mu)\dd \mu.\label{eq:flux}
\end{equation}

\subsection{Numerical computation of the concentration field: the Legendre Polynomial Spectral Method (LPSM)}
\label{sec:LPSM}

 In this section, we outline the numerical method used to solve for the advection-diffusion problem, Eqs.~\eqref{eq:advdiff},  and compute the nutrient uptake for a given stroke $\boldsymbol\alpha$. \change{The} method is based on the expansion of the different fields using  Legendre polynomials in $\mu$ \change{and generalizes the approach presented in Ref.~\cite{magar2003} to the entire stroke space.}

The nutrient concentration $c(r,\mu)$ is decomposed onto Legendre polynomials  as
\begin{equation}\label{eq:leg_exp}
c(r,\mu)=\sum_{m=0}^\infty C_m(r)L_m(\mu).
\end{equation}
Substituting Eq.~\eqref{eq:leg_exp} into Eqs.~\eqref{eq:advdiff} leads after projection on the $p$-th Legendre polynomial ($p\geq 0$) to a system of coupled ODEs in $r$
\begin{align}
\Pe\sum_{m=0}^\infty\sum_{n=1}^\infty\alpha_n\left(\right.&\left.A_{mnp}\totd{C_m}{r}\psi_n+B_{mnp}C_m\totd{\psi_n}{r}\right)\label{eq:sys1}\\
&=r^2\totd{^2C_p}{r^2}+2r\totd{C_p}{r}-p(p+1)C_p,\nonumber\\
C_p(1)&=\delta_{p1},\\ 
C_p(\infty)&=0,
\end{align}
where the functions $\psi_n(r)$ are defined in Eq.~\eqref{eq:stream_n} and $A_{mnp}$ and $B_{mnp}$ are third order scalar tensors defined in Appendix \ref{sec:AB}. The relative nutrient flux is then obtained simply as
\begin{equation}\label{eq:flux_lpsm}
J=-\totd{C_0}{r}(r=1).
\end{equation}

In the numerical simulations, the summations in Eq.~\eqref{eq:sys1} are truncated at a finite number $N$ of squirming modes to describe the swimming stroke ($1\leq n\leq N$), and $M$ Legendre polynomial modes are used to describe the azimuthal variations of $c$ ($0\leq m\leq M-1$). Adapting the technique used in Ref.~\cite{magar2003}, the system of ODEs in $r$ is discretized on a stretched grid obtained by mapping as $r=\ee^{\phi(\xi)}$ a uniformly-spaced grid of $N_r$ points in $\xi$. The choice of an exponential stretching allows to cover both  far-field and near-field concentrations. The function $\phi$ is a third-order polynomial in $\xi$ such that a fixed fraction of the total number of points are contained within the expected concentration boundary layer at intermediate and high $\Pe$ number. The discretized system \eqref{eq:sys1} can then be rewritten as 
\begin{equation}
\mathbf{H}\cdot\mathbf{C}=\mathbf{R},
\end{equation}
where $\mathbf{C}$ is a $N_r\times M$ vector containing the values of $C_m(r_j)$ on the different grid points, and $\mathbf{H}$ is a $M\times M$ block-matrix, each block being tridiagonal of size $N_r\times N_r$. The block structure of $\mathbf{H}$ is tightly-banded: $\mathbf{H}$ is diagonal if $N=0$, tridiagonal if $N=1$, pentadiagonal if $N=2$, etc. The contribution to the right-hand side $\mathbf{R}$ arises from the non-homogeneous boundary condition on the swimmer surface for the first mode $C_0(r)$. This large linear system is solved using a direct block-Gaussian elimination technique taking advantage of the sparse structure of $\mathbf{H}$.

\subsection{Results}
\label{sec:results_concentration}

\begin{figure*}
\begin{center}
\includegraphics[width=18cm]{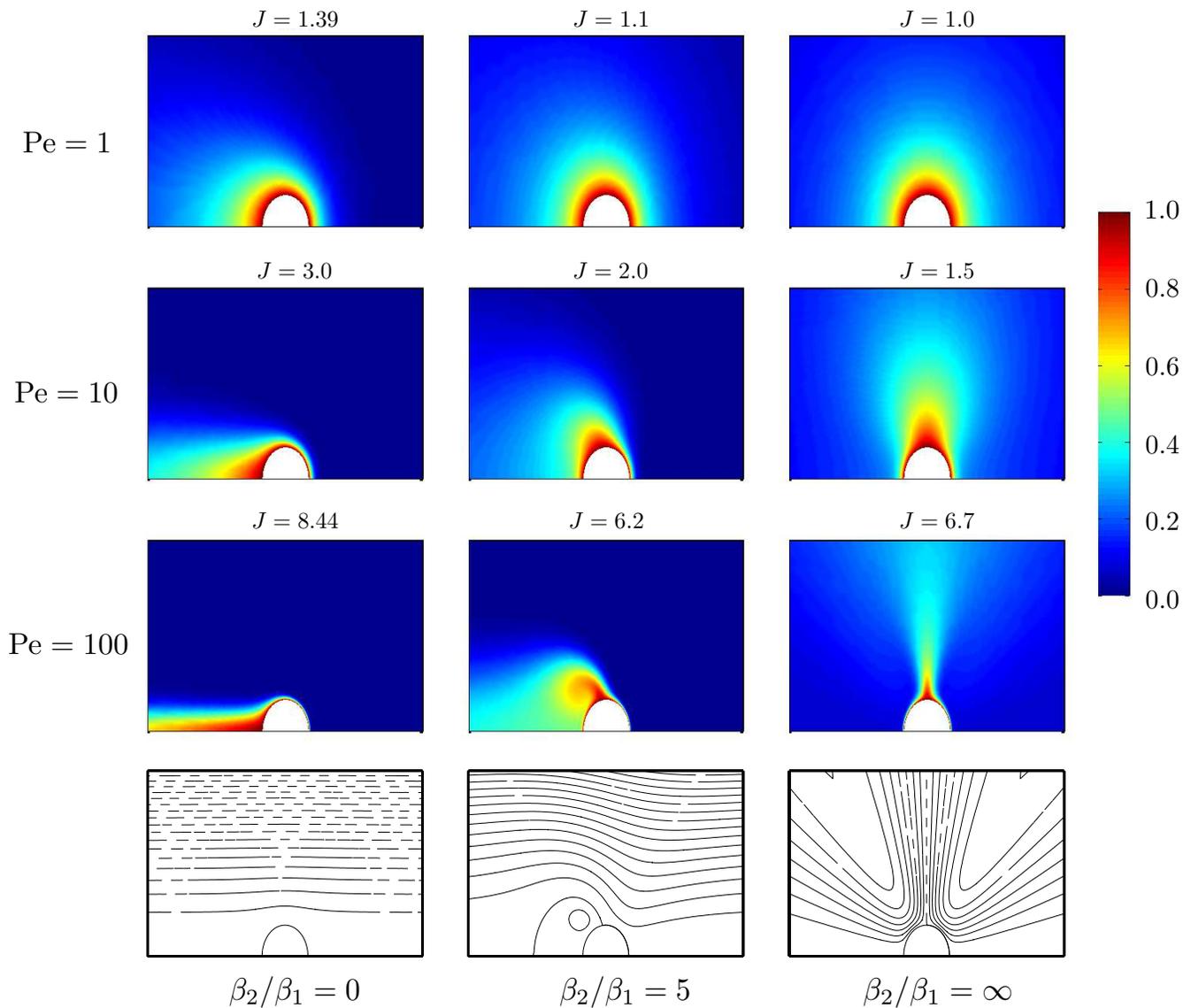}
\caption{(Color online) Nutrient concentration around the swimmer for $\Pe=1$, $10$ and $100$ (from top to bottom) and $\beta_2/\beta_1$=$0$, $5$ and $\infty$ (from left to right), all the other $\beta_j$ being taken equal to zero. Far from the swimmer $c=0$, while $c=1$  at the swimmer surface. The dimensionless nutrient flux $J$ is quoted for each case. On the bottom row, the streamlines are displayed for each stroke.} \label{fig:scalar_concentration}
\end{center}
\end{figure*}

The Legendre Polynomial Spectral Method (LPSM) presented in the previous section is now used to compute, for different values of $\Pee$, the nutrient concentration around a squirmer for  simple steady swimming strokes including only the first two squirming modes. In Fig.~\ref{fig:scalar_concentration}, the concentration field is shown for three different values of $\Pee$ ($1$, $10$ and $100$) and three different swimming strokes: the pure treadmill ($\beta_n=\delta_{n1}$), a combination of modes $1$ and $2$, and a pure mode-$2$ stroke. Note that in the latter case, the organism is not swimming. The corresponding relative nutrient uptake $J$ is given for each case, and the streamlines (independent of the value of $\Pee$) are also shown for each stroke.

At low $\Pee$ (typically $\Pe\leq 1$), the concentration distribution is close to isotropic, and only a few Legendre modes are necessary to compute $c(r,\mu)$ accurately. The far-field behavior is reached rapidly, so $\phi(\xi_\textrm{max})=8$ is sufficient with $N_r\sim 80$--$100$ to achieve errors of at most $0.1$--$0.5$\% on the nutrient uptake. Note from Fig.~\ref{fig:scalar_concentration} that the nutrient concentration is not very sensitive to the swimming stroke, resulting in similar relative nutrient uptake $J$. In that regime, the typical diffusion time is much shorter than the advective time, resulting in the homogenization of the concentration field and a weak front-back asymmetry along the swimming direction.

As the value of $\Pee$ is increased, the concentration distribution develops a stronger angular asymmetry as a nutrient-depleted wake ($c$ close to $1$) develops in the region ``behind" the organism. Molecular diffusion is not rapid  enough  to homogenize the sharper advection-induced gradients as both processes now act on the same time scale. \change{This} applies for swimming ($\beta_1\neq 0$) as well as non-swimming strokes ($\beta_1=0$). 

For $\Pe\gg 1$, a boundary layer develops for the nutrient concentration in the region where the flow impinges on the swimmer surface. In the regions where the radial flow leaves the swimmer surface (wake of the treadmill swimmer or upward direction for the pure stresslet swimmer) a nutrient-depleted region forms where molecular diffusion processes do not have the time to smooth out the sharp concentration gradients induced by the velocity field. Numerically, more Legendre modes are required (typically $M\sim 100$ for $\Pe\sim 10$ up to $M\sim 400$ for $\Pe\sim 400$), and one needs to extend the $r$-grid further in the far-field (up to $\phi(\xi_\textrm{max})\sim 18$--$20$ for the highest values of $\Pee$ considered) and increase its resolution (up to $N_r\sim 400$ for the highest values of $\Pee$ considered).

Figure \ref{fig:scalar_concentration} shows that for a given stroke the relative nutrient uptake, $J$, is an increasing function of $\Pee$, emphasizing the systematic benefit of the swimming or stirring motion on the feeding process. For a fixed $\Pee$ ({\it i.e.}~constant energy cost), \change{it also  shows} that the treadmill swimmer always performs better than the two other strokes considered. The pure treadmill and pure mode-$2$ strokes share the existence of a sharp nutrient-depleted ejection zone. However, one notices easily that the gradients at the surface of the organism are stronger in the former case due to the swimming motion of the organism toward a nutrient-rich zone. Swimming appears therefore, in these preliminary results, to be a more efficient process than simple stirring. The intermediate stroke is characterized by a recirculation region that tends to create at high $\Pee$ a zone of homogenized nutrient concentration in the wake of the swimmer, thereby reducing the radial gradients in that region as well as the nutrient uptake.

\begin{figure}[h]
\begin{center}
\includegraphics[width=8.5cm]{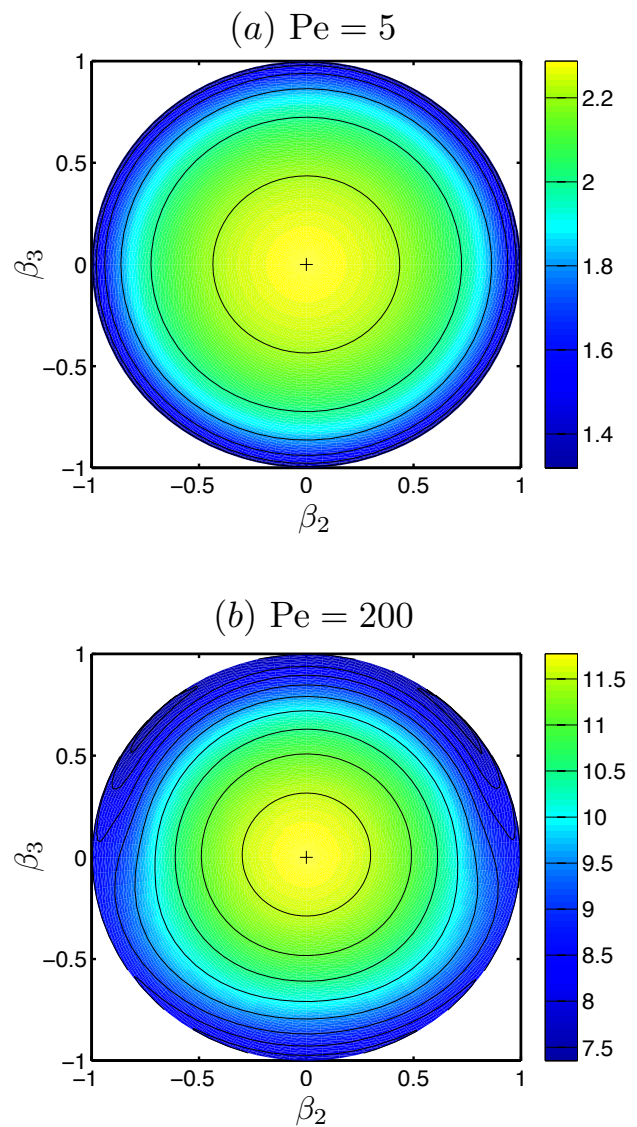}
\caption{(Color online) Variations of the relative nutrient flux, $J$, within the $(\beta_2,\beta_3)$-plane for (a) $\Pe=5$ and (b) $\Pe=200$ ($\beta_1$ is adjusted so that $\sum\beta_j^2=1$). Nutrient flux isolines are also shown in black for clarity and correspond to the values indicated in the colorbars. The crosses indicate the position of the treadmill swimmer in the $(\beta_2,\beta_3)$-plane.}\label{fig:scan_J}
\end{center}
\end{figure}

To confirm this optimality of the treadmill swimmer, \change{Fig.~\ref{fig:scan_J} shows} the value of the nutrient uptake for all possible swimming strokes obtained with only the first three squirming modes  $(\beta_1,\beta_2,\beta_3)$  (this consists of a two-dimensional space because of the constant energy constraint, Eq.~\eqref{eq:energy_constraint}). 
We see in Fig.~\ref{fig:scan_J} that, within this three-parameter family of flow profiles, the optimal feeding swimmer is located around $\beta_2\sim\beta_3\sim 0$, corresponding to the treadmill swimmer.

\section{Optimal feeding by a steady squirmer}
\label{sec:optimal_feeding}

The results of the previous section suggest that the treadmill swimmer ($\beta_n=\delta_{n1}$) is a feeding optimum at all P\'eclet numbers. In this section this result \change{is confirmed} by considering the formal optimization problem of the relative nutrient flux $J$ with respect to the swimming stroke $\alphab$. We start in Sec.~\ref{sec:adjoint} by  presenting the general framework of adjoint-based optimization for a swimmer of time-independent arbitrary shape $\Sc$ prescribing tangential surface velocities $\ub^S$ on its boundaries. Although the results presented in the remainder of the paper correspond to the simplest geometry (a sphere), this framework can be applied to organisms of arbitrary shapes and is of interest for a large variety of advection-diffusion problems. We then focus on the particular squirmer configuration in Sec.~\ref{sec:gradient_squirmer}, and present our optimization results in Sec.~\ref{sec:results_opt}. We show that the optimal feeding stroke is essentially the same as the optimal swimming one, a result true for all values of the P\'eclet number. \change{These numerical results are also confirmed using asymptotic analysis predictions  outlined in Appendices~\ref{sec:asymptotic_smallPe} and \ref{sec:asymptotic_largePe}.}

\subsection{Nutrient uptake gradient for a general swimmer}
\label{sec:adjoint}

To derive the optimal swimmer, the gradient of $J$ with respect to the swimming stroke must be mathematically determined. This gradient indicates the changes to make in the swimming stroke in order to increase $J$, leading to a natural computational implementation of the optimization search.

The gradient is obtained using variational analysis as in Ref.~\cite{michelin2010c}.  Considering a small variation $\delta\ub^S$ of the swimming stroke, and the corresponding change in the flow velocity field $\delta\ub=\mathcal{L}\cdot\delta\ub^S$ (see Eq.~\ref{eq:lin_op}), the resulting change $\delta\Phi$ is given by
\begin{equation}
\delta\Phi=-\frac{1}{\Pe}\int_\Sc\pard{(\delta c)}{n}\dd S,
\end{equation}
where $\nb$ is the outward normal unit vector to the surface of the swimmer and, at leading order, $\delta c$ is the solution of 
\begin{subeqnarray}
\Pe\ub\cdot\grad \delta c-\nabla^2 \delta c&=&-\Pe\delta\ub\cdot\grad c\slabel{eq:nutrientdelta_eq},\\
\delta c=0 \textrm{\,   for\,   } &\xb&\in\Sc\textrm{\,  and\,  }\xb\rightarrow\infty\slabel{eq:nutrientdelta_bc}.
\end{subeqnarray}
Multiplying Eq.~\eqref{eq:nutrientdelta_eq} by a test function $g$ and integrating over the entire fluid domain $V_f$, \change{one obtains after integration by part} that, at leading order, 
\begin{equation}\label{eq:gradient_phi}
\delta\Phi=-\int_{V_f} c(\mathcal{L}\cdot\delta\ub^S)\cdot\grad g\,\dd V
,\end{equation}
provided that the function $g$ satisfies the adjoint equation:
\begin{subeqnarray}
\Pe\ub\cdot\grad g&=&-\nabla^2 g,\\
g&=&1 \textrm{\,   for\,   } \xb\in\Sc,\\
 g&\rightarrow& 0 \textrm{ \, for\,  }\xb\rightarrow\infty\slabel{eq:adjoint_bc}.
\end{subeqnarray}
Equation \eqref{eq:gradient_phi} defines the gradient of the absolute nutrient uptake with respect to the swimming stroke. Since $\Phi_0$ does not depend on the imposed surface velocity, the gradient of the relative nutrient uptake $J$ is obtained similarly. Note that the adjoint field $g$ satisfies the same advection-diffusion equation as the original passive scalar after replacing $\Pee$ by $-\Pee$ (or alternatively $\ub^S$ by $-\ub^S$), so the same analytical or numerical methods can be implemented to solve for both fields.

\subsection{Nutrient uptake optimization for a squirmer}
\label{sec:gradient_squirmer}
In the particular case of a squirmer, the gradient of the relative nutrient uptake $J$ with respect to the swimming stroke $\alphab$ is obtained from Eq.~\eqref{eq:gradient_phi} as
\begin{equation}\label{eq:gradient_stream}
\pard{J}{\alpha_n}=-\frac{\Pe}{2}\int_1^\infty\int_{-1}^1c(r,\mu)\left[\pard{\Psi_n}{r}\pard{g}{\mu}-\pard{\Psi_n}{\mu}\pard{g}{r}\right]\dd \mu\,\dd r.
\end{equation}
Numerically, both the concentration and adjoint fields are determined for a given swimming stroke $\alphab$ using the method outlined in Sec.~\ref{sec:LPSM}. The relative nutrient uptake $J$ is then obtained from $C_0(r)$ as in Eq.~\eqref{eq:flux_lpsm}. Its gradient with respect to $\alpha_n$ is computed as
\begin{align}\label{eq:grad_lpsm}
\pard{J}{\alpha_n}=-\Pe\sum_{m=0}^\infty\sum_{p=0}^\infty&\left[\frac{A_{mnp}}{2p+1}\int_1^\infty C_p\psi_n\totd{G_m}{r}\dd r\right.\nonumber\\
&\left.+\frac{B_{mnp}}{2p+1}\int_1^\infty C_p\totd{\psi_n}{r}G_m\dd r\right],
\end{align}
where the functions $G_m(r)$ are defined in analogy with $C_m(r)$ from the adjoint field $g(r,\mu)$. All the above integrals are well defined, taking into account the far-field behavior of $c$ and $g$ and the definitions of $A_{mnp}$ and $B_{mnp}$.

In the following, the optimal steady swimming stroke for a given energy consumption is determined ({\it i.e.}~the optimal $\alphab$ or $\betab$ at given $\Pee$). Starting from a random initial condition $\betab^{(0)}$ on the unit hypersphere, the following steepest ascent algorithm is \change{applied}:
\begin{enumerate}
\item{At step $k$, for a given stroke $\betab^{(k)}$, the LPSM is used to solve for the concentration field $c$ and its adjoint $g$. The value of the corresponding nutrient flux $J^{(k)}$ is also computed from Eq.~\eqref{eq:flux_lpsm}.}
\item{From Eqs.~\eqref{eq:beta} and \eqref{eq:grad_lpsm}, the gradient $\grad_\beta J$ of the relative nutrient flux is computed.}
\item{At fixed $\Pe$, $\betab^{(k)}\cdot\betab^{(k)}=1$ and the gradient tangential to the unit hypersphere is obtained by projection
\begin{equation}\label{eq:proj_grad}
\grad_\parallel J=\grad_\beta J-\left(\betab^{(k)}\cdot\grad_\beta J\right)\betab^{(k)}.
\end{equation}
}
\item{$\grad_\parallel J$ defines the steepest ascent direction on the unit hypersphere in $\beta$-space and the next iteration is carried at a new guess for the optimal $\betab$
\begin{equation}\label{eq:march}
\betab^{(k+1)}=\frac{\betab^{(k)}+s\grad_\parallel J}{|\betab^{(k)}+s\grad_\parallel J|},
\end{equation}
until convergence is reached to a local maximum when it is not possible to find a new guess with $J^{(k+1)}>J^{(k)}$ using this procedure, even in the limit $s\rightarrow 0$.}
\end{enumerate}

\subsection{Results}\label{sec:results_opt}

\subsubsection{Optimal squirmer for various $\Pe$ numbers}
\label{sec:optim_results}

The preliminary results obtained in Sec.~\ref{sec:results_concentration} suggest that the treadmill swimming stroke corresponds to the optimal feeding mechanism at all P\'eclet numbers. \change{This result is confirmed here using the numerical optimization techniques outlined above.} An arbitrary stroke is characterized by an infinite number of coefficients $\beta_n$; for numerical purpose, this description must be truncated to the first $N$ squirming modes, thereby exploring a reduced stroke-space. The results of the stroke optimization are presented below for the cases $N=3$ and $N=8$. \change{Computations performed for larger values of $N$ led essentially to the same optimal strokes and feeding rates.}

For given values of $N$ and $0.01\leq \Pe\leq 300$, several optimization runs were performed starting with different random initial strokes. In each run, a rapid convergence was observed toward an optimal stroke, only marginally different from the treadmill swimmer (pure mode~1).

\begin{figure}[t]
\begin{center}
\includegraphics[width=8.5cm]{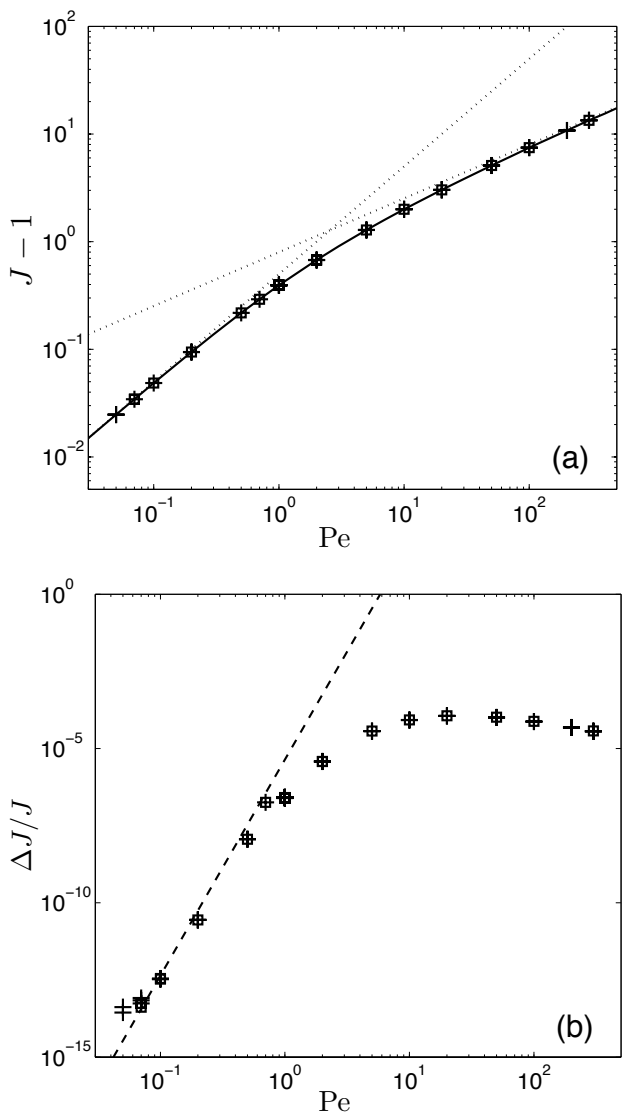}
\caption{(a) Optimal stroke-induced nutrient flux $J-1$ and (b) Relative difference in nutrient flux, $\Delta J/J$, between the optimal swimmer and the treadmill swimmer as functions of the P\'eclet number, $\Pee$. Numerical results of the optimization procedure are presented for $N=3$ (crosses) and $N=8$ (squares). Several sets of calculations were performed for each value of $\Pee$ and $N$. In (a), the solid line corresponds to the treadmill swimmer. In (a) and (b), the dashed and dotted lines correspond to the asymptotic results for the treadmill swimmer at $\Pe\ll 1$ and $\Pe\gg 1$ obtained in \change{Appendices~\ref{sec:asymptotic_smallPe} and \ref{sec:asymptotic_largePe}.} }\label{fig:J_optim}
\end{center}
\end{figure}

The variation  of the optimal feeding rate with the P\'eclet number,  $\Pee$, is shown in Fig.~\ref{fig:J_optim}(a) and emphasizes the strong gain in feeding rate associated with the performance of the swimming and/or stirring motion. As $J=1$ corresponds to the case of a rigid sphere ($\Pe=0$), the quantity plotted on Fig.~\ref{fig:J_optim}(a), $J-1$, is a measure of the excess rate of feeding induced by the surface motion. Figure~\ref{fig:J_optim}(a) also compares the results of the computational optimization procedure for two different values of $N$ with the feeding rate obtained for the treadmill swimmer. \change{The main observation is} that although the rate of feeding is strongly dependent on the value of the P\'eclet number, the numerical optimal is  undistinguishable at this scale from that of the treadmill swimmer for all values of the P\'eclet number. The asymptotic scalings for the treadmill nutrient uptake $J_\textrm{treadmill}$ are obtained in \change{Appendices~\ref{sec:asymptotic_smallPe} and \ref{sec:asymptotic_largePe}} (see also Ref.~\cite{magar2003})
\begin{align}\label{eq:Jtread_scaling}
J_\textrm{treadmill}&\sim 1+\frac{\Pe}{2}\textrm{ \,  for\,    } \Pe\ll 1, \\
 J_\textrm{treadmill}&\sim\sqrt{\frac{2\Pe}{\pi}}\textrm{\,   for\,    } \Pe\gg 1,
\end{align}
\change{and show an excellent agreement with the numerical results (Fig.~\ref{fig:J_optim}a). }

The relative difference in nutrient flux, $\Delta J/J$,  between the numerical optimal and that of the  treadmill swimmer is \change{shown} in Fig.~\ref{fig:J_optim}(b). We see that it is always small -- below $10^{-3}$ -- across the investigated range of P\'eclet numbers \change{and is maximum around $\Pe\approx 10$.} A clear power-law scaling can be observed at low $\Pee$ for $\Delta J/J$;  for $\Pe\leq 1$, this power-law behavior is in excellent agreement with \change{the predictions of the asymptotic analysis (see \change{Appendix~\ref{sec:asymptotic_smallPe}}):
\begin{equation}\label{eq:reldJ_asympt}
\frac{\Delta J}{J}\sim\left(\frac{2161}{1034880}\right)^2\,\Pee^7\approx 4.36\,10^{-6}\,\Pee^7.
\end{equation}
 }
 
As a side note,  the computational  results above are presented only for $\Pe\geq 0.05$. Below this value, the optimization algorithm is unable to find optimal strokes performing better than the treadmill swimmer. This does not rule out the existence of a different optimum, but indicates that this optimum differs from the treadmill swimmer by an amount smaller than the round-off error of our computations. 

\begin{figure}
\begin{center}
\includegraphics[width=8.5cm]{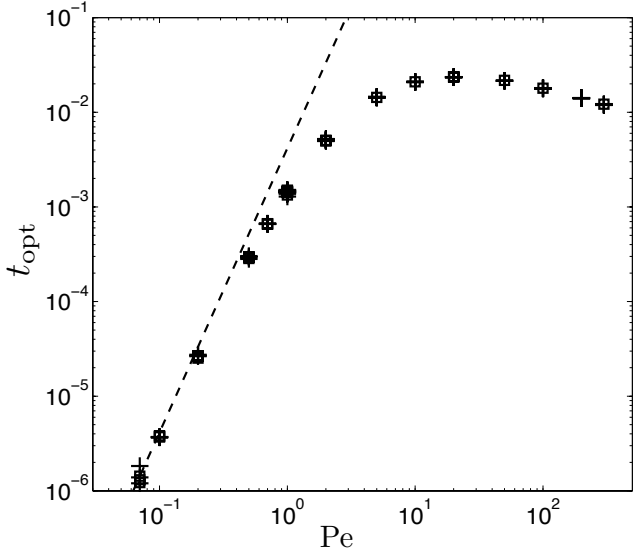}
\caption{Dependence on the P\'eclet number,  $\Pee$, of the orientation angle \change{in $\beta$-space, $t_\textrm{opt}=\cos^{-1}\beta_1$, of the optimal swimming stroke}. As in Fig.~\ref{fig:J_optim}, results are presented when the optimization is performed on $N=3$ modes (crosses) and $N=8$ modes (square). The dashed line corresponds to the prediction of the asymptotic analysis at $\Pe\ll 1$ obtained in \change{Appendix~\ref{sec:asymptotic_smallPe}}.}\label{fig:solid_angle}
\end{center}
\end{figure}
We now turn to the description of the optimal swimming stroke. \change{In the $\beta$-space where the stroke is represented by a point on the unit hyper-sphere, the difference to the treadmill swimmer is measured using the angle $t_\textrm{opt}=\cos^{-1}\beta_1$ between the optimal stroke $\betab$ and the treadmill stroke.  The fraction of the stroke energy cost not dedicated to the swimming velocity, {\it i.e.}~the energy of the non-swimming modes, is $\sin^2 t_\textrm{opt}$ and is directly related to the swimming efficiency $\eta$ of the stroke \cite{michelin2010c}: $t_\textrm{opt}\sim\sqrt{1-2\eta}$. The variation of $t_\textrm{opt}$ with $\Pee$ is shown on Fig.~\ref{fig:solid_angle}. For all $\Pe$, the difference between the treadmill and optimal strokes remains small, with $t_\textrm{opt}\leq 0.02$, corresponding to an energy in the non-swimming modes accounting to less than $0.04\%$ of the total energy cost. However, this small difference depends strongly on $\Pe$ (see Fig.~\ref{fig:solid_angle}). At low $\Pe$, $t_\textrm{opt}$ scales as $\Pe^3$, and for $\Pee\leq 1$ the numerical results are in excellent agreement with the predictions of the asymptotic analysis (see \change{Appendix~\ref{sec:asymptotic_smallPe}}):
\begin{equation}\label{eq:t_asympt}
t_\textrm{opt}\sim\frac{2161}{517440}\Pee^3\approx 0.00418\,\Pee^3.
\end{equation}
For $\Pe\gg 1$, $t_\textrm{opt}$ scales as $t_\textrm{opt}\sim \Pe^{-1/3}$  (Fig.~\ref{fig:solid_angle}). Note that the similarity in shape of Figs.~\ref{fig:J_optim}(b) and \ref{fig:solid_angle} is a direct result of $t_\textrm{opt}\ll 1$, as explained below.}

\subsubsection{Gradient near the treadmill}
\label{sec:gradient}

The optimal feeding squirmer is essentially\change{, but not exactly,} identical to the treadmill swimmer. \change{Therefore, its properties and feeding rate are expected to be determined by the nutrient flux gradient $\grad J$ in the stroke space, evaluated at the treadmill. For $t\ll 1$, the swimming stroke 
\begin{equation}
\betab=\cos t\,\betab_1+\sin t\,\betab_\parallel,
\end{equation}
is a perturbation of the stroke from the pure treadmill, $\betab_1$ in the direction $\betab_\parallel$, such that $\betab_\parallel\cdot\betab_1=0$ ({\it i.e.}~non-swimming stroke).} 
Then, the nutrient flux can be expanded near $\betab_1$ as
\begin{equation}
J=J_1+t\left(\pard{J}{\beta_\parallel}\right)_1+\frac{t^2}{2}\left[\left(\pard{^2J}{\beta_\parallel^2}\right)_1-\left(\pard{J}{\beta_1}\right)_1\right]+O(t^3),
\end{equation}
\change{where derivatives with a $1$ subscript are evaluated at the treadmill. The nutrient flux is therefore maximum for the treadmill if and only if:}
\begin{equation}\label{eq:optim_cond}
\left(\pard{J}{\beta_\parallel}\right)_1=0 \quad\textrm{     and     }\quad\left(\pard{^2J}{\beta_\parallel^2}\right)_1<\left(\pard{J}{\beta_1}\right)_1\cdot
\end{equation}
More generally, the optimal value of $t$ and corresponding flux \change{are} given at leading order by 
\begin{subeqnarray}
t_\textrm{opt}&\sim&\frac{\left(\partial J/\partial\beta_\parallel\right)_1}{\left(\partial J/\partial \beta_1\right)_1-\left(\partial^2J/\partial\beta_\parallel^2\right)_1},\\
 \frac{\Delta J}{J}&\sim&\frac{\left[\left(\partial J/\partial\beta_\parallel\right)_1\right]^2}{2J_1\left[\left(\partial J/\partial \beta_1\right)_1-\left(\partial^2J/\partial\beta_\parallel^2\right)_1\right]}\cdot
\end{subeqnarray}

\change{These results emphasize the critical role of the nutrient flux gradient $\partial J/\partial\beta_j$ in the localization of the optimal feeding stroke with respect to the treadmill. Integrating Eq.~\eqref{eq:gradient_stream} by part, the gradient can be rewritten as
\begin{align}
\pard{J}{\alpha_n}=-\frac{\Pe}{4}&\int_1^\infty\int_{-1}^1\left[\pard{\Psi_n}{r}\left(c\pard{g}{\mu}-g\pard{c}{\mu}\right)\right.\nonumber\\
&\left.+\pard{\Psi_n}{\mu}\left(g\pard{c}{r}-c\pard{g}{r}\right)\right]\dd\mu\,\dd r.\label{eq:gradient_sym}
\end{align}
Using the parity properties in $\mu$ of $\Psi_n$, one easily obtains that for the treadmill, $g(r,\mu)=c(r,-\mu)$, and consequently
\begin{equation}
\forall p\geq 1,\quad \left(\pard{J}{\beta_{2p}}\right)_1=0.
\end{equation}
}
\begin{figure}
\begin{center}
\includegraphics[width=8.5cm]{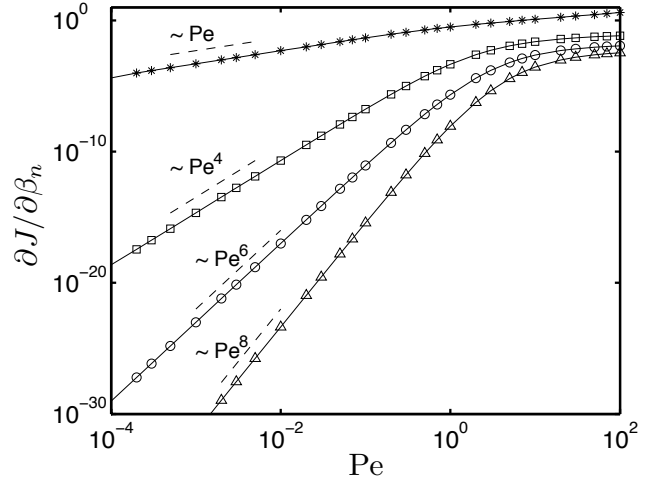}
\caption{Dependence with the P\'eclet number, $\Pee$, of the nutrient flux gradient $\partial J/\partial\beta_n$ with respect to the first four odd modes $n=1$ (stars), $n=3$ (squares), $n=5$ (circles) and $n=7$ (triangles) 
and evaluated at the treadmill (the even mode gradients are equal to zero by symmetry). The power law dependence of each component is indicated by a dashed line.}\label{fig:gradient_treadmill}
\end{center}
\end{figure}

\change{The variation with $\Pee$ of the first four \change{odd (and non-trivially zero)} gradient components of $J$ at the treadmill is plotted in Fig.~\ref{fig:gradient_treadmill}. Clear scalings are identified for $\Pe\ll 1$ and we obtain by regression \footnote{Note that the gradients considered here are absolute gradients, computed before the projection on the hypersphere as detailed in Eq.~\eqref{eq:proj_grad}.}
\begin{equation}\label{eq:gradient_trend}
\left(\pard{J}{\beta_1}\right)_1\approx 0.49\,\Pe,\qquad
 \left(\pard{J}{\beta_{3}}\right)_1\approx 0.0020\, \Pee^4,
\end{equation}
and more generally 
\begin{equation}
\left(\pard{J}{\beta_{2p+1}}\right)_1=O(\Pee^{2p+2})\qquad \textrm{for   }p\geq 1.
\end{equation}
These results are in good agreement with the predictions of the asymptotic analysis at low $\Pee$ (see Appendix~\ref{sec:asymptotic_smallPe})
\begin{equation}
\pard{J}{\beta_1}\sim\frac{\Pee}{2},\qquad\pard{J}{\beta_3}\sim\frac{2161}{1034880}\Pee^4\approx 0.0021\,\Pee^4.
\end{equation}

Figure~\ref{fig:gradient_treadmill} also shows that the gradient along the third mode always dominates by at least one order of magnitude the gradients in the other directions, for all $\Pee$. The difference between the optimal and treadmill strokes is therefore expected to be dominated by the third squirming mode which is confirmed by the fact that the results presented in Section \ref{sec:optim_results} are only marginally modified between $N=3$ and $N=8$. 

For large $\Pee$, $\partial J/\partial \beta_{2p+1}\ll \partial J/\partial \beta_1$, which is consistent with the asymptotic result that the feeding rate only depends on $\beta_1$ at leading order (see Appendix~\ref{sec:asymptotic_largePe}).
}

\section{Discussion}\label{sec:discussion}

In this paper, the steady spherical squirmer model \change{was used} to determine optimal feeding strategies at zero Reynolds number. For a nutrient following an advection-diffusion equation, we showed computationally and theoretically that, for a fixed amount of energy dissipated in the fluid,  the optimal feeding mechanism is essentially equivalent to the optimal swimming problem, and its solution maximizes the swimming velocity.

Perhaps surprisingly, the result that optimal feeding is optimal swimming does not  depend on the value of the P\'eclet number, which \change{is} confirmed  by asymptotic analysis. At low P\'eclet, the improvement in feeding rate as compared to quiescent fluid environment (pure nutrient diffusion) is, as expected, small and, it increases as $\Pee$. This linear scaling arises from the proportionality between the  gain in nutrient uptake and  the volume swept  by the swimming organism, which itself is proportional to its surface area times its swimming speed.  
In the high P\'eclet regime, the development of concentration boundary layers means that the  volume swept by the swimming organism decreases, and the relative nutrient uptake shows a slower increase with $\Pee$ than linear.

One interesting feature of the optimal feeding (equivalently, optimal swimming) solution is that it is vorticity free. This surface treadmill solution corresponds indeed to the only surface distribution of velocity which leads to potential flow Stokesian swimming. This result could very well be a simple consequence of our emphasis on  energy cost, as the presence of vorticity always increases the rate of energy dissipation \cite{stone1996}.

 \change{Note that the occurrence of a  $\Pee$-independent optimal feeding stroke in our simulations is  reminiscent of results  on optimal tracer mixing obtained for all $\Pee$  using flows directed from sources to sinks \cite{shaw2007
 }. Here, the optimal stroke corresponds to the swimmer (a sink) maximizing its velocity toward the sources of nutrients in the far-field.}

One of the major assumptions of our modeling approach is the restriction of the study to steady surface motion. In the case of our work on locomotion optimization \cite{michelin2010c}, we showed that although the treadmill swimmer is itself not physical (due to the non-periodicity of the trajectories), the unsteady optimum  was found to be   a superposition of the treadmill solution with periodic shock-like recovery strokes where material elements on the organism surface were brought back to their initial position. We conjecture that the same will be true in the case of feeding, and that the solution to the optimal feeding for periodic surface motion will be a combination of the optimal steady (treadmill) with regularization to enforce periodicity at a rate allowed by the energetic constraints.  Ongoing work in this direction, technically more complex as it requires solving for the spatio-temporal evolution of both the concentration field and the adjoint field, will be reported in the future.

\begin{acknowledgements}
This work was supported in part by the US National Science Foundation (grant CBET-0746285 to E. L.).
\end{acknowledgements}

\appendix
\section{Definition of the $A_{mnp}$ and $B_{mnp}$ tensors}
\label{sec:AB}
The coefficients $A_{mnp}$ and $B_{mnp}$ used in Section \ref{sec:LPSM} are defined in terms of the Legendre polynomials as follow:
\begin{align}
A_{mnp}=&\frac{(2p+1)(2n+1)}{2}\int_{-1}^1L_m\,L_n\,L_p\,\dd\mu,\label{eq:defA}\\
B_{mnp}=&\frac{(2p+1)(2n+1)}{2n(n+1)}\int_{-1}^1(1-\mu^2)L'_m\,L'_n\,L_p\,\dd \mu.\label{eq:defB}
\end{align}
They are easily computed using 
\begin{equation}
A_{m0p}=\delta_{mp},\qquad B_{m0p}=0.
\end{equation}
and the following recursive relations for $n\geq 1$
\begin{align}
A_{mnp}=&\frac{2n+1}{n}\left[-\frac{n-1}{2n-3}A_{m,n-2,p}+\frac{m+1}{2m+1}A_{m+1,n-1,p}\right.\nonumber\\
&\left.+\frac{m}{2m+1}A_{m-1,n-1,p}\right],\\
B_{mnp}=&\frac{2n+1}{n(n+1)}\left[\frac{(n-2)(n-1)}{2n-3}B_{m,n-2,p}\right.\nonumber\\
&\left.+\frac{m(m+1)}{2m+1}\left(A_{m-1,n-1,p}-A_{m+1,n-1,p}\right)\right].
\end{align}

\section{Asymptotic analyis: optimal feeding for $\Pee\ll 1$}\label{sec:asymptotic_smallPe}
\change{In this appendix, we focus on the treadmill stroke $\beta_j=\delta_{j1}$, and determine the concentration field $c$, nutrient flux $J$ and nutrient flux gradient for $\Pe\ll 1$ using asymptotic analysis.}
\subsection{Concentration field around the treadmill}

For $\Pe\ll 1$, $c(r,\mu)$ is sought in the form of a regular perturbation expansion in $\Pe$:
\begin{equation}\label{eq:outer_pert}
c(r,\mu)=\sum_{p=0}^\infty \Pee^p c_p(r,\mu),
\end{equation}
\change{with $c_0=1/r$, the rigid sphere ($\Pee=0$) solution.} However, this expansion is not uniformly valid over the entire fluid domain and one must consider a boundary layer at infinity for $\Pe>0$ \cite{acrivos1962,magar2003}. In the near-field (outer solution), $c$ must satisfy the advection-diffusion equation
\begin{align}
\frac{1}{r^2}\left[\pard{}{r}\right.&\left.\left(r^2\pard{c}{r}\right)+\pard{}{\mu}\left((1-\mu^2)\pard{c}{\mu}\right)\right]\nonumber\\
=&-\Pe\left[\mu\left(1-\frac{1}{r^3}\right)\pard{c}{r}+\frac{1-\mu^2}{r}\left(1+\frac{1}{2r^3}\right)\pard{c}{\mu}\right],\label{eq:outer_eq}
\end{align}
as well as $c=1$ on the swimmer surface. In the boundary layer $\cin(\rho,\mu)=c(r,\mu)$, with $\rho=\Pe r$, must instead satisfy the boundary-layer equation as
\begin{align}
\frac{1}{\rho^2}\left[\pard{}{\rho}\right.&\left.\left(\rho^2\pard{\cin}{\rho}\right)+\pard{}{\mu}\left((1-\mu^2)\pard{\cin}{\mu}\right)\right]+\mu\pard{\cin}{\rho}\nonumber\\
&+\left(\frac{1-\mu^2}{\rho}\right)\pard{\cin}{\mu}=\frac{\Pee^3}{\rho^3}\left[\mu\pard{\cin}{\rho}-\frac{1-\mu^2}{2\rho}\pard{\cin}{\mu}\right],\label{eq:inner_eq}
\end{align}
as well as $\cin\rightarrow 0$ for $\rho\rightarrow\infty$.

\change{Both $c$ and $\cin$ are sought as regular perturbation series in $\Pe$. Using Matched Asymptotic Expansion \cite{bender1978}, both solutions are computed up to order $O(\Pe^p)$ ($p=1,2,3$) and integration constants at each order are obtained by identifying the two solutions  up to terms $O(\Pe^p,\Pe^{p-1}/r,...,1/r^p)$ over a matching region $\Pee^{-p/(p+1)}\ll r\ll\Pee^{-1}$.} 

The final solution valid up to $O(\Pe^4)$ is given by
\begin{equation}\label{eq:out_sol}
c(r,\mu)=\sum_{p=0}^3\Pe^p\sum_{q=1}^p c_p^{q}(r)L_q(\mu),
\end{equation}
valid for   $1\leq r\ll \Pee^{-1}$ and 
\begin{align}
\cin(\rho,\mu)=\left\{\Pe+\frac{\Pee^2}{2}\right.&\left.+\Pee^3\left[\frac{17}{80}+\frac{3\mu}{8}\left(1+\frac{2}{\rho}\right)\right]\right\}\nonumber\\
&\times\frac{1}{\rho}\textrm{exp}\left[-\frac{(1+\mu)\rho}{2}\right],
\label{eq:inn_sol}
\end{align}
valid for   $\Pee^{1/(p+1)}\ll \rho$, where the functions $c_p^q(r)$ are defined in Appendix \ref{sec:small_Pe_app}. 
Using Eq.~\eqref{eq:flux_lpsm}, the nutrient flux is then obtained as
\begin{equation}
\label{eq:flux_smallPe}
J=1+\frac{\Pe}{2}-\frac{13\,\Pe^2}{80}+\frac{7\,\Pe^3}{80}+O(\Pe^4).
\end{equation}

\subsection{Gradient computation}
\change{Using the previous expansion and Eq.~\eqref{eq:gradient_stream}, one can compute $\partial J/\partial \beta_n$ at the treadmill. Using the front-back symmetry of the treadmill velocity field, the asymptotic expansion of the adjoint field is obtained as $g(r,\mu)=c(r,-\mu)$ and $\gin(r,\mu)=\cin(r,-\mu)$. Splitting the integral in $r$ in Eq.~\eqref{eq:gradient_stream} between inner and outer regions, one obtains
\begin{align}
\pard{J}{\alpha_n}&=-\frac{\Pe}{2}\left(I_\textrm{int}+I_\textrm{BL}\right)\label{eq:gradient_split}\\
I_\textrm{int}&=\int_1^{\small{\Pee}^{-7/8}} \left\{\totd{\psi_n}{r}\mathscr{F}_n\left[c\pard{g}{\mu}\right]+\psi_n\mathscr{F}_n^*\left[c\pard{g}{r}\right]\right\}\dd r \label{eq:gradient_Iint} \\
I_\textrm{BL}&=\int_{\small{\Pee^{1/8}}}^\infty\left\{\totd{\psi_n}{\rho}\mathscr{F}_n\left[\cin\pard{\gin}{\mu}\right]+\psi_n\mathscr{F}_n^*\left[\cin\pard{\gin}{\rho}\right]\right\}\dd \rho\label{eq:gradient_IBL}
\end{align}
where $\mathscr{F}_n$ and $\mathscr{F}_n^*$ are projection operators on the Legendre polynomials
\begin{align}
\mathscr{F}_n[f](r)&=\frac{2n+1}{n(n+1)}\int_{-1}^1f(r,\mu)(1-\mu^2)L_n'(\mu)\dd\mu,\\
 \mathscr{F}^*_n[f](r)&=(2n+1)\int_{-1}^1f(r,\mu)L_n(\mu)\dd\mu.
\end{align}
For $n=3$,  the integral in Eq.~\eqref{eq:gradient_IBL} is at least $O(\Pee^4)$. Using Eq.~\eqref{eq:out_sol} and the definition of $c_p^q$ in Appendix~\ref{sec:small_Pe_app}, the gradient with respect to the third mode is then computed as 
\begin{equation}
\pard{J}{\beta_3}=\frac{2161}{1034880}\Pee^4+O(\Pee^5).
\end{equation}
Following a similar approach, the gradient with respect to $\beta_1$ is computed as
\begin{equation}
\pard{J}{\beta_1}=\frac{\Pee}{2}+O(\Pee^2).\label{eq:gradb1}
\end{equation}

Note that a similar but longer approach consists in computing the gradients directly from the expansion of $c$ for an arbitrary combination of two modes \cite{magar2003}.  This calculation, omitted here for clarity, also provides the second derivative $\partial^2 J/\partial \beta_3^2$ evaluated at the treadmill:
\begin{equation}
\pard{^2J}{\beta_3^2}=\frac{27}{7840}\Pee^2+O(\Pee^3).
\end{equation}
}

Using the results of Sec.~\ref{sec:gradient}, $J$ has a maximum in the $(\beta_1,\beta_3)$-space at $\beta_3^\textrm{opt}$ corresponding to a relative increase $\Delta J/J$ of the nutrient flux:
\begin{subeqnarray}\label{eq:low_Pe_result}
\beta_3^\textrm{opt}&\sim&\frac{2161}{517440}\Pee^3\approx 0.00418\,\Pee^3,\\
 \frac{\Delta J}{J}&\sim& \left(\frac{2161}{1034880}\right)^2\Pee^7\approx 4.36\,10^{-6}\Pee^7.
\end{subeqnarray}

\section{Asymptotic analysis: optimal feeding at $\Pe\gg1$}\label{sec:asymptotic_largePe}
\change{As shown in Fig.~\ref{fig:scalar_concentration}, the feeding problem at $\Pee\gg 1$ is characterized by the formation of a boundary layer in the concentration distribution near the squirmer's surface, whose thickness scales as $\Pe^{-1/2}$ due to the balance between tangential advection and radial diffusion near the swimmer's surface. Generalizing the analysis in Ref.~\cite{magar2003} to arbitrary strokes, Eq.~\eqref{eq:nutrient} becomes at leading order in $1/\sqrt{\Pe}$, 
\begin{equation}
\pard{^2c}{R^2}=\zeta'(\mu)R\pard{c}{R}-\zeta(\mu)\pard{c}{\mu}, \label{eq:BL_eq}
\end{equation}
with $R=\sqrt{\Pe}(r-1)$ and $\zeta$ the axial component of the tangential surface velocity:
\begin{equation}
\zeta(\mu)=\sqrt{1-\mu^2}\,u_\theta^S(\mu).
\end{equation}}
A self-similar solution is sought for the previous equation in terms of the variable $\eta=R/g(\mu)$, where $g(\mu)$ represents the boundary layer thickness. Equation \eqref{eq:BL_eq} then becomes
\begin{equation}
\pard{^2c}{\eta^2}=\eta\pard{c}{\eta}\left\{\zeta(\mu)g(\mu)g'(\mu)+g(\mu)^2\zeta'(\mu)\right\}.\label{eq:BL_eq_eta}
\end{equation}
Provided that
\begin{equation}
\zeta(\mu)g(\mu)g'(\mu)+g(\mu)^2\zeta'(\mu)=-2,\label{eq:self_sim}
\end{equation}
a self-similar solution compatible with the boundary conditions Eq.~\eqref{eq:nutrient_bc1}-\eqref{eq:nutrient_bc2} exists, given by
\begin{equation}
c(R,\mu)=\frac{2}{\sqrt{\pi}}\int_{R/g(\mu)}^\infty\ee^{-\eta^2}\dd\eta.\label{eq:self_sim_sol}
\end{equation}

Equation \eqref{eq:self_sim} can be solved for $g(\mu)$ with the additional constraint that the boundary layer thickness $g(\mu)$ is finite at $\mu=1$
\begin{equation}
g(\mu)=\frac{2}{\zeta(\mu)}\sqrt{\int_\mu^1\zeta(t)\dd t}\,.\label{eq:BL_g}
\end{equation}
If the surface velocity $u_\theta^S$ is positive everywhere ({\it i.e.}~there are no recirculation regions), then $g(\mu)$ remains finite for all $\mu>-1$, and the boundary layer solution above is valid over the entire surface of the squirmer. This condition is satisfied by the treadmill swimmer, and in some vicinity of it. At $\mu=-1$, $g(\mu)=\infty$ and the boundary layer separates into the wake observed on Fig.~\ref{fig:scalar_concentration}.

The nutrient flux at the surface of the squirmer is then obtained from Eqs.~\eqref{eq:flux} and \eqref{eq:self_sim_sol} as
\begin{equation}
J=\sqrt{\frac{\Pe}{\pi}}\int_{-1}^1\frac{\dd\mu}{g(\mu)}+O(1),
\end{equation}
and can be computed exactly from Eq.~\eqref{eq:BL_g} as
\begin{equation}\label{finalJ_highPe}
J=\sqrt{\frac{\Pe}{\pi}\int_{-1}^1\zeta(\mu)\dd\mu}+O(1)=\sqrt{\frac{2\beta_1\Pe}{\pi}}+O(1).
\end{equation}
\change{One observes that at leading order $J$ depends on $\beta_1$ only.}

\section{Definition of the functions $c_p^q(r)$}
\label{sec:small_Pe_app}
The functions $c_p^q(r)$ in Eq.~\eqref{eq:out_sol} are given by
\begin{align}
c_1^0=&\frac{1}{2}\left(\frac{1}{r}-1\right)\\
c_1^1=&-\frac{1}{2}+\frac{3}{4r^2}-\frac{1}{4r^3}\\
c_2^0=&\frac{r}{6}-\frac{1}{4}+\frac{7}{80r}+\frac{1}{24r^2}-\frac{1}{16r^4}+\frac{1}{60r^5}\\
c_2^1=&\frac{r}{4}-\frac{1}{4}+\frac{1}{8r^2}-\frac{1}{8r^3}\\
c_2^2=&\frac{r}{12}-\frac{1}{4r}+\frac{5}{24r^2}+\frac{3}{56r^3}-\frac{1}{8r^4}+\frac{5}{168r^5}\\
c_3^0=&-\frac{r^2}{24}+\frac{r}{12}-\frac{17}{60}+\frac{11}{240r}+\frac{1}{48r^2}-\frac{1}{96r^4}+\frac{1}{120r^5}\\
c_3^1=&-\frac{3r^2}{40}-\frac{r}{8}-\frac{23}{160}-\frac{3}{40r}+\frac{527}{1120r^2}-\frac{11}{320r^3}\nonumber\\
&-\frac{3}{112r^4}-\frac{3}{560r^5}+\frac{3}{160r^6}-\frac{9}{2240r^7}\\
c_3^2=&-\frac{r^2}{24}+\frac{r}{24}-\frac{1}{12r}+\frac{5}{48r^2}-\frac{5}{336r^3}-\frac{1}{48r^4}+\frac{5}{336r^5}\\
c_3^3=&-\frac{r^2}{120}+\frac{3}{80}-\frac{1}{20r}-\frac{9}{560r^2}+\frac{3}{40r^3}-\frac{9}{224r^4}\nonumber\\
&-\frac{9}{1120r^5}+\frac{1}{80r^6}-\frac{1}{420r^7}
\end{align}


\end{document}